\documentclass[12pt,journal]{IEEEtran}
\usepackage{cite}

\usepackage{graphicx}
\usepackage{subfigure}
\usepackage{epstopdf}
\usepackage{multicol}
\usepackage{algorithm}
\usepackage{algorithmic}
\usepackage{amsthm,amsmath,amsfonts}
\usepackage{mathrsfs}
\usepackage{courier}

\usepackage{bm}
\usepackage{subfigure}

\usepackage{color}

\def\BibTeX{{\rm B\kern-.05em{\sc i\kern-.025em b}\kern-.08em
    T\kern-.1667em\lower.7ex\hbox{E}\kern-.125emX}}

\begin{document}
%
\title{CoMP-Enabled RAN Slicing for Tactile Internet}
%
%
%
\author{Peng Yang, Xing Xi, Tony Q. S. Quek, and Hyundong Shin
}

\maketitle

\begin{abstract}
Tactile Internet (TI) enables the omnipresence and exchange of tactile experiences across the globe via the ultra-reliable and ultra-responsive connectivity.
This article argues for coordinated multi-point (CoMP) enabled radio access network (RAN) slicing as an efficient solution that satisfies the stringent reliable and responsive connectivity requirements for supporting tactile applications.
This article presents the emerging challenges when accommodating CoMP-enabled RAN slicing in the TI ecosystem and expounds on the functional split of a CoMP-enabled RAN.
The implementation prototype of CoMP-enabled RAN slicing for TI is also elaborated when the tactile applications coexist with other vertical applications.
Finally, this article studies a use case of enabling TI-included application multiplexing as an example of CoMP-enabled RAN slicing for TI.
\end{abstract}


%
\IEEEpeerreviewmaketitle

\section{Introduction}
\IEEEPARstart{T}{he} Tactile Internet (TI), poised to shift networks from content transmission to skill-set transmission, is envisioned to revolutionize most segments of our society \cite{simsek20165g}.
To realize the grand envision, TI is expected to deliver remote control and physical haptic information in real time via the ultra-reliable and ultra-responsive (i.e., the end-to-end (E2E) latency on the order of one millisecond) network connectivity \cite{aijaz2016realizing}.
Owing to the stringent reliability and response requirements, one might be persuaded to construct a network separately designed for E2E control and physical haptic communications.
This scheme, however, is infeasible in terms of the capital (CAPEX) and operational (OPEX) expenditures.
To improve the cost efficiency of providing E2E tactile services, TI should be established under the existing and extending physical network infrastructure.
Furthermore, there is a consensus that existing networks must evolve in the direction of simultaneously supporting multiple vertical applications (e.g., haptic communications, massive machine type communications (mMTC), and enhanced mobile broadband (eMBB) \cite{series2015imt}) with completely different requirements on reliability, latency, throughput, and availability \cite{simsek20165g}.

\subsection{Definition of CoMP-enabled RAN slicing}
Radio access network (RAN) slicing is a crucial concept of enabling the existing network infrastructure to be efficiently shared among diverse vertical applications \cite{rost2017network}.
The concept of RAN slicing is defined for running multiple logical or virtual networks as independent business operations on a common RAN.
Each network slice indicates an independent logical network tailored to accommodate the quality-of-service (QoS) requirement of a particular application.
Besides, to achieve the hard connectivity requirements, some advanced radio access technologies (RATs) such as massive MIMO, coordinated multi-point (CoMP) transmission, full-duplex are required to be explored \cite{aijaz2016realizing}.
Especially, CoMP transmission, creating spatial diversity with redundant communication paths, has been extensively considered as one of the most significant RATs that fulfill the hard-bound reliability requirement for tactile applications \cite{simsek20165g,Coordinated31GPP}.
Hence, the concept of CoMP-enabled RAN slicing is defined as the partition of a RAN exploring the CoMP transmission technique (called CoMP-enabled RAN) into multiple virtual networks.
Nevertheless, adapting CoMP-enabled RAN slicing to the TI ecosystem brings many potential challenges, in part owing to the difficulties in abstracting the CoMP-enabled RAN into multiple slices, as described below.

\subsection{Design challenges}
\textbf{Effective network resource allocation:}
Network slicing will quickly run into the limited availability of physical network resources (e.g., transmit power and spectrum).
This limitation is aggravated when network slice resources are allocated based on tactile traffic with fully predictable arrival time and amount.
If the arrival time and the number of tactile packets cannot be accurately predicted (e.g., in telesurgery) \cite{simsek20165g}, effectively allocating network resources to guarantee ultra-reliable control and haptic information transmission will be challenging. Besides, to satisfy the stringent low-latency requirement, tactile packets must be immediately transported once generated. Tactile packets will be blocked and even dropped if not enough resources are allocated for the slice carrying tactile applications (or TI slice for brevity), in turn reducing the reliability.

\textbf{Multi-timescale issue:}
Network slicing operations (e.g., activate and release network slices) will be performed in the timescale of minutes to hours to keep in sync with the slicing of upper layers. In upper-layer slicing processes, some protocols (e.g., RAN protocol stacks) and functions (e.g., radio resource control (R$^2$C)) will be activated and configured, the operations of which are time-consuming \cite{rost2017network}. Nevertheless, wireless channels may change on the order of a millisecond to seconds. Remarkably, the time interval during which the wireless channel remains unchanged may be much shorter than the RAN slicing operation duration. Then, optimally conducting time-consuming RAN slicing operations in rapidly varying wireless channels, which is the so-called multi-timescale issue of network slicing, should be tackled and is challenging.

\textbf{Coordination latency reduction:}
CoMP transmission acting as a spatial diversity technique that has been standardized \cite{Coordinated31GPP} can significantly improve transmission reliability under the low-latency constraint \cite{simsek20165g}.
Further, it can be used to avoid retransmissions through the message (e.g., payload, command, and channel state information (CSI)) exchange among all remote radio heads (RRHs) for coordination.
However, the message exchange results in non-negligible latency, especially in scenarios with low-latency requirements.
How to reduce coordination latency in CoMP transmission is a big challenge.

\subsection{Related work and contributions}
\subsubsection{Related work}
The concept of network slicing has been extensively investigated in the literature, wherein a dedicated portion of RAN elements is fully reserved for specific services.
Nevertheless, with TI's advent, network slicing for TI has evolved in more agile network frameworks, aiming to attain immersive haptic interactions and enable varying levels of human-in/out-of-the loop services.
For instance, in \cite{promwongsa2020ensuring}, the problem of placing virtualized network functions (VNFs) in a VNF chain, which constituted network slices for serving tactile applications, and routing tactile traffic through the chain to fulfill the hard low-latency mandate was studied.
Based on radio-aware software-defined networking (SDN) and NFV techniques, the work in \cite{simsek20165g} developed a flexible network architecture of creating different network slices to serve TI-included multiple applications on common programmable physical infrastructure.
In \cite{ksentini2018providing}, a two-level medium access control (MAC) scheduling framework for reducing transmission scheduling latency in network slices, which were constructed to enable TI-involved application multiplexing over a shared RAN, was developed.
Besides, an elastic cloud-based RAN (CRAN) architecture providing multi-connectivity for tactile users via dynamically splitting network slices was proposed in \cite{shafigh2017dynamic}.

\subsubsection{Our contributions}
In this article, we set forth the fundamental pillars for the exploration of the CoMP-enabled RAN slicing concept in TI in detail based on the 3GPP recommended centralized unit (CU) and distributed unit (DU) structure for the next generation RAN architecture \cite{Study38GPP}.
Specific focus is put on the basic architectural principles for incorporating CoMP-enabled RAN slicing into the TI ecosystem. In this regard, this article underlines the crucial domains of TI architecture, putting emphasis on the functional split of a CoMP-enabled RAN for TI. Besides, this article elaborates on the implementation prototype of CoMP-enabled RAN slicing that enables the multiplexing of TI-involved multiple vertical applications.

\begin{figure*}[!t]
\centering
\includegraphics[width=3.6in]{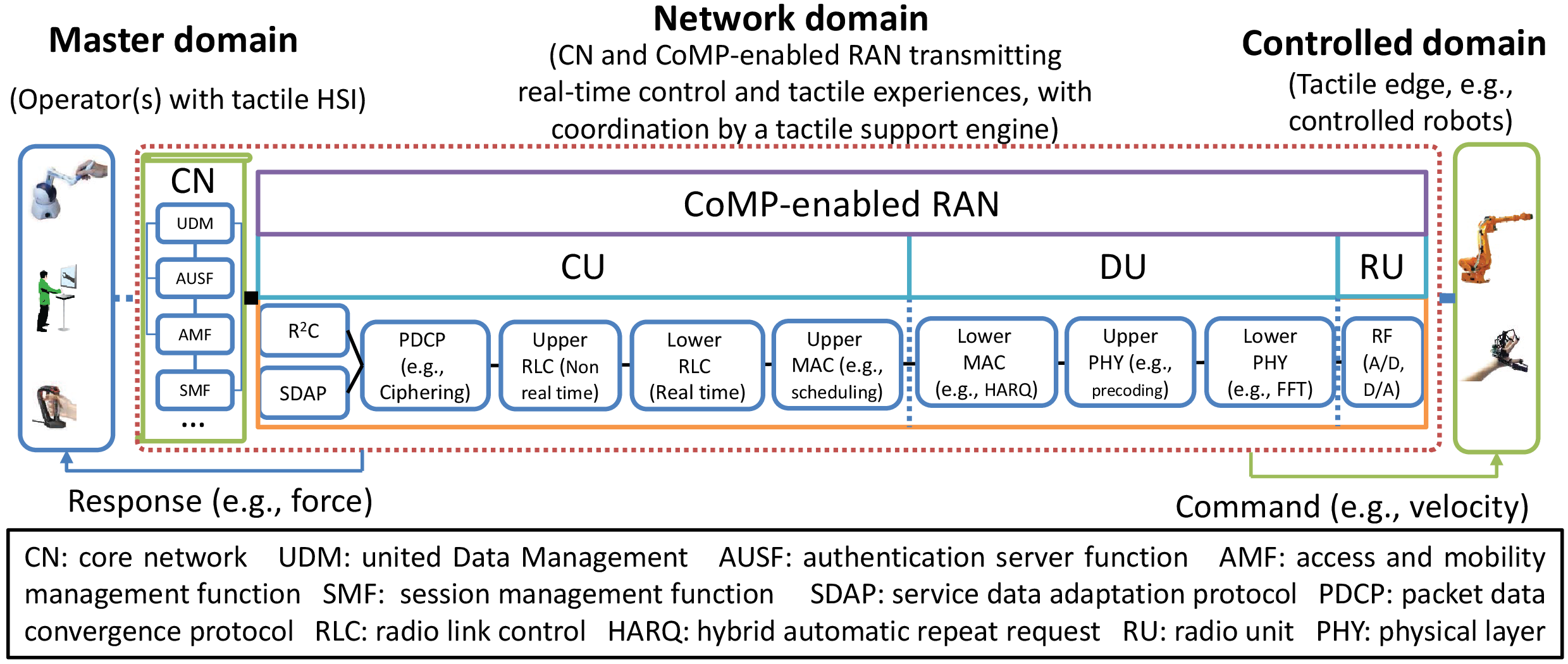}
\caption{TI architecture and functional split of a CoMP-enabled RAN.}
\label{fig_network_domain}
\end{figure*}
\section{TI architecture and functional split}
TI offers the medium for delivering touch and actuation in real time. The haptic sense occurs bilaterally; that is, haptic information is sensed by imposing an action on an unknown environment and feeling the environment by a reaction force or a distortion.
In this way, TI realizes the haptic control via the Internet, which is quite different from the conventional Internet. For the conventional Internet, the feedback can only be audio/visual information.
The exploration on the TI architecture is a crucial problem of enabling TI. Figure \ref{fig_network_domain} depicts an E2E architecture of TI, which is divided into three different domains; that is, a master domain, a controlled domain, and a network domain \cite{simsek20165g}.

\subsection{Master domain}
The master domain comprises a tactile operator (human) and a human-system interface (HSI) that can be a haptic device. The HSI is used to convert the tactile operator's inputs to haptic inputs via tactile coding techniques.
Through operating (e.g., touch, feel, and manipulate) the HSI, the tactile operator can control objects in the controlled domain.
Except for outputting haptic information, the master domain has the provisioning of visual/audio sensory feedback owing to multi-dimensional characteristics of human tactile perception. By integrating diverse sensory modalities, the human perception performance can be significantly improved, and thereby the haptic and non-haptic feedback
greatly help the tactile operator realize the immersive control and steering of the controlled domain.

\subsection{Controlled domain}
The controlled domain mainly includes a controlled robot(s) responding to tactile operator's commands to control various objects in the unknown environment coordinately. The robot reports both cutaneous and kinaesthetic data and leverages codec to report haptic feedback to the tactile operator.
Thus, the energy between the master and controlled domains is exchanged through command and haptic (and audio/visual) feedback signal delivery.

\subsection{Network domain}
The network domain constructs a network architecture for bilateral communications between the master and controlled domains. It aims to provide the tactile operator with an immersive experience in the remote environment.
To this aim, TI needs the ultra-reliable and ultra-responsive network connectivity, as mentioned at the beginning.
The network architecture comprising a CN and a CoMP-enabled RAN and the correspondingly functional split, as shown in Figure \ref{fig_network_domain}, is expected to be designed to fulfill the critical connectivity requirements.

The CN is in charge of security processing, defining the access levels of services available to diverse applications, and handling connection and mobility management tasks. Accordingly, the CN comprises some dedicated hardware or virtualized modules (e.g., AUSF, AMF, and home subscribe server (HSS)).
Significant functions of the CoMP-enabled RAN comprise ultra-reliable and low-latency packet delivery, cost-efficient QoS aware scheduling, and efficient resource management.
Hence, the CoMP-enabled RAN is split into a CU, a DU pool, and many RUs. This paradigm of functional split also leads to cost reduction and dynamic scalability via virtualization.
Particularly, RUs (or RRHs) comprise radio-frequency circuitry indicating that RUs' functionalities are implemented as physical network functions (PNFs).
Other functionalities distributed over the CU and the DU pool can be virtually implemented as VNFs.
The CU domain contains higher-layer functionalities, and the DU pool includes lower-layer functionalities.
For example, high-level MAC scheduling decisions (i.e., CoMP) are made in the CU while the time-critical MAC processing (e.g., HARQ) resides in the DU pool.

There is a consensus that future networks (including the CoMP-enabled RAN) must be flexibly designed to be efficiently shared by diverse vertical applications from both academia and industry.
The efficient network sharing can be achieved by network slicing with diverse network slices being assigned to serve different types of vertical applications.
We next present the implementation prototype of CoMP-enabled RAN slicing for TI with the coexistence of other vertical applications and study a use case of application multiplexing.

\section{Implementation prototype and case study}
\subsection{Implementation prototype of CoMP-enabled RAN slicing for TI}
\begin{figure}[!t]
\centering
\includegraphics[width=2.7in]{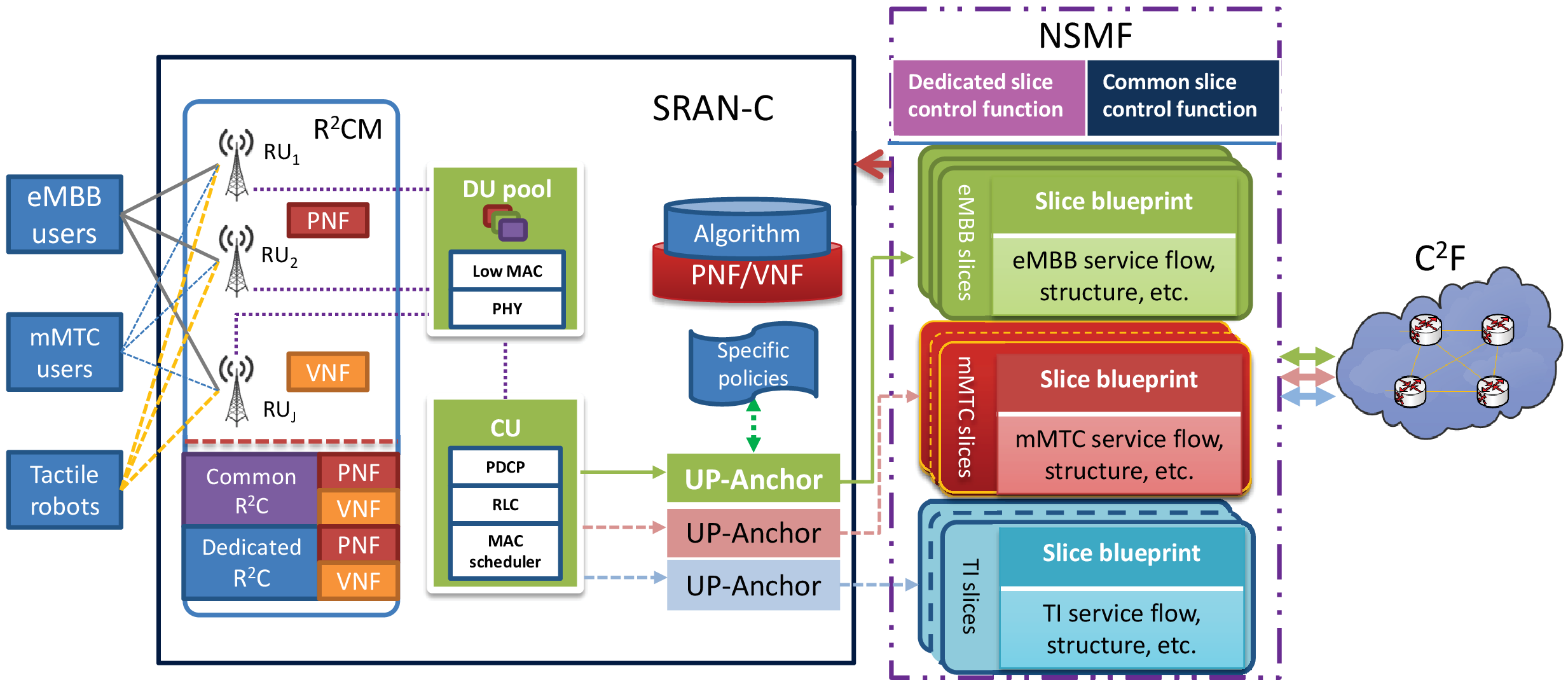}
\caption{Schematic of the implementation prototype of CoMP-enabled RAN slicing for TI.}
\label{fig_RAN_slicing_archi}
\end{figure}
Figure \ref{fig_RAN_slicing_archi} illustrates how to implement the CoMP-enabled RAN slicing for TI, the prototype of which
comprises three major components \cite{yang2021multicast}:
{Software-defined CoMP RAN coordinator (SRAN-C): }calculate and update accepted network slices;
{Network slice management function (NSMF): }a virtualized function used to manage as well as control network slices;
{CN control function (C$^2$F): }aim at configuring the CN based on slices' QoS requirements.

The SRAN-C module comprises five parts: radio resource control and management (R$^2$CM), DU pool, CU, user plane anchor (UP-Anchor), and algorithm.
It is time-consuming to create and configure network slices.
If some slices are created, then R$^2$CM needs to configure RAN protocol stacks based on network slices' QoS requirements.
For instance, some schemes like shortened orthogonal frequency division multiplexing (OFDM) symbol duration, lower frame error rates, and CoMP transmission are desired to be explored for TI slices.
Inevitably, the exploration of the CoMP transmission technique results in undesired coordination latency. Yet, we can resort to the scalable numerology (e.g., the elastic number of OFDM symbols in a transmit time interval (TTI) and subcarrier spacing) to bridge the coordination latency to some extent.
Additionally, common/dedicated R$^2$C functions in R$^2$CM will be activated for application-specific control of radio resources based on VNF/PNF \cite{rost2017network}.
The DU pool is introduced to implement PHY collaboration facilitating the CoMP deployment.
MAC scheduler aims at scheduling traffic according to network conditions to relieve network congestion.
Depending on particular applications, CU will configure and tailor corresponding UP protocol stacks, for instance, for TI slices, Internet protocol (IP) and header compression techniques will be disabled, and the RLC function will be transparently enabled in CU \cite{rost2017network}.
Due to the requirement for high capacity (data rates) between the CU and the DUs, optical fiber communications are ideal or non-constrained fronthaul. Some lower-layer protocols like the open base station architecture initiative (OBSAI) and the common public radio interface (CPRI) \cite{ericsson2015common} with a goal of unifying the fronthaul interface also need to be adopted.
UP-Anchor aims to distribute traffic on the basis of the configured network slice policy and conduct encryption
with slice-specific keys as well.
For instance,
policy requirements for reliability and security need to be identified on priority for TI slices.
Besides, according to network slice requests and available resources, the algorithm repetitively calculates, updates, and reconfigures accepted network slices to obtain the maximum slice utility. The SRAN-C module also realizes the RAN-CN mapping that is done by leveraging some configuration protocols (e.g., FlexRAN) \cite{ksentini2017toward}.

The NSMF deployed in physical/virtualized network infrastructure is in charge of creating, maintaining, activating, configuring, and releasing network slices during their life cycles. Through dedicated/common network slice control functions, NSMF generates a network slice blueprint (i.e., a template) for every network slice. The network slice blueprint includes the structure, configuration, control signals, and service flows of a network slice, which are used to instantiate and control the network slice. A network slice instance comprises a collection of network functions and resources configured to fulfill an application's QoS requirement.

C$^2$F will interpret a network slice blueprint when the network slice request arrives. Specifically, it will configure the network based on the interpreted blueprint and identify the optimal paths and servers to deploy VNFs such that the slice's QoS requirement can be satisfied.
It also reserves several data centers for supporting network slices. Every data center comprises some servers (e.g., HSS) to provide VNFs' services (e.g., identity, independent subscription and session) for a network slice. Data centers are connected through high-capacity backhaul and can offer services separately or jointly.

\subsection{Case study of CoMP-enabled RAN slicing: TI-involved application multiplexing}
Based on the implementation prototype, we provide a case study of CoMP-enabled RAN slicing for TI, mMTC, and eMBB multiplexing to show its potentials in application multiplexing.
Recall that the solution to the challenging issues of effective network resource allocation and multi-timescale are inevitable when investigating the CoMP-enabled RAN slicing.
Therefore, we formulate the CoMP-enabled RAN slicing for TI, mMTC, and eMBB multiplexing as a multi-timescale resource optimization problem with a goal of maximizing the overall slice utilities.
For each type of applications, its slice utility is defined to reflect users' QoS mandates in the application and RUs' total transmit power consumption.
Particularly, we use the E2E latency, codeword error decoding probability, and packet blocking probability to characterize the QoS requirements of tactile robots \cite{yang2021multicast}. The achievable data rates and random access (RA) success probabilities of users are used to measure eMBB and mMTC users' QoS requirements, respectively \cite{yang2020ran}.
Besides, in this problem, the system bandwidth allocated to different network slices is optimized every ten minutes, and the RU transmit power is optimized every ten seconds.
This problem is constrained by i) total system bandwidth; ii) the maximum RU transmit power;
iii) QoS requirements of tactile robots and eMBB and mMTC users.

The formulated multi-timescale optimization problem is confirmed to be mixed-integer non-convex. By exploring an alternating direction method of multipliers (ADMM) \cite{yang2021multicast}, we propose a novel iterative resource allocation (IRA-ADMM) algorithm to solve the problem.
Specifically, by exploring sample average approximations (SAA), IRA-ADMM converts the multi-timescale problem to multiple mixed-integer non-convex subproblems of single timescale. To make the subproblems tractable, an iterative optimization scheme is firstly explored to handle the mixed-integer issue. Next, a semidefinite relaxation (SDR) scheme \cite{yang2021multicast} is applied to transform the non-convex subproblems into convex ones. The solutions (i.e., the bandwidth allocated to each network slice and RU transmit power) to the single timescale subproblems can then be obtained.
Finally, IRA-ADMM aggregates the achieved slice bandwidth in the single timescale into the multi-timescale one using the ADMM.

Next, we design a simulation to verify the effectiveness of IRA-ADMM and compare IRA-ADMM with an iterative resource allocation (IRA) algorithm. Differing from IRA-ADMM, IRA does not recover the multi-timescale slice bandwidth from obtained single-timescale slice bandwidth results. We consider a one km$^2$ communication area wherein all tactile robots and eMBB and mMTC users are randomly and uniformly distributed, and three RUs are uniformly deployed to serve these robots and users coordinately.
There are two TI slices, three eMBB slices, and three mMTC slices. The first TI slice includes three tactile robots with a latency requirement of one millisecond. The second TI slice contains five robots with a latency requirement of two milliseconds. The codeword error decoding probability and packet blocking probability requirements of all tactile robots are set to be $10^{-5}$ and $2\times 10^{-8}$, respectively. Tactile packets are generated by a Poisson distribution \cite{li20185g}.
The number of users in three eMBB slices is four, six, and eight with the achievable data rate requirements of 6 Mb/s, 4 Mb/s, and 2 Mb/s, respectively. There are $600$ mMTC users in each mMTC slice with each user's RA success probability requirement being $0.5$. Other parameter settings are as follows: noise power is $-110$ dBm, total system bandwidth is $4$ MHz, the maximum transmit power of each RU is $1$ W, each RU is equipped with two antennas, all robots and users are single antenna-equipped, the length of a tactile packet is $160$ bits \cite{yang2021multicast,yang2020ran}.
\begin{figure}[!t]
\centering
\includegraphics[width=2.5in]{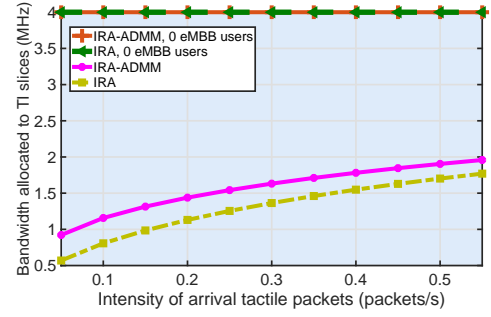}
\caption{Bandwidth allocated to TI slices vs. intensity of arrival {tactile} packets.}
\label{fig_TI_Bandwidth_intensity}
\end{figure}

Figure \ref{fig_TI_Bandwidth_intensity} shows the relationship between the bandwidth allocated to TI slices and the tactile packet arrival rate when considering the TI and eMBB multiplexing.
We observe that both IRA-ADMM and IRA will allocate the total system bandwidth to TI slices to guarantee the QoS performance of TI slices when terminating eMBB services. When tactile and eMBB applications share the network infrastructure, the network bandwidth allocated to TI slices monotonously increases with an increasing tactile packet arrival rate. This observation indicates that the proposed algorithm can effectively solve the resource allocation issue.
\begin{figure}[!t]
\centering
\includegraphics[width=2.5in]{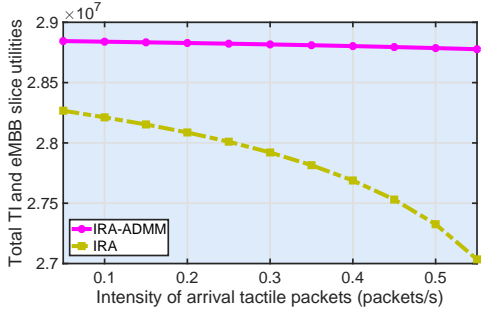}
\caption{Total TI and eMBB slice utilities vs. intensity of arrival {tactile} packets.}
\label{fig_TI_eMBB_intensity}
\end{figure}

Figure \ref{fig_TI_eMBB_intensity} plots the trend of the total TI and eMBB slice utilities over different tactile packet arrival rates.
We observe that the obtained total TI and eMBB slice utilities of IRA-ADMM and IRA monotonously decrease with an increasing tactile packet arrival rate. Besides, IRA-ADMM achieves greater total slice utilities than IRA.
As an increasing tactile packet arrival rate results in a decreasing amount of network bandwidth allocated to eMBB slices, greater RU transmit power should be configured to ensure QoS requirements of eMBB slices. Thus, a small eMBB slice utility incurring the reduction of the total slice utilities is obtained.
IRA-ADMM obtains greater TI slice utility than IRA because IRA-ADMM suggests to allocate more bandwidth to TI slices, as shown in Figure \ref{fig_TI_Bandwidth_intensity}.
\begin{figure}[!t]
\centering
\includegraphics[width=2.5in]{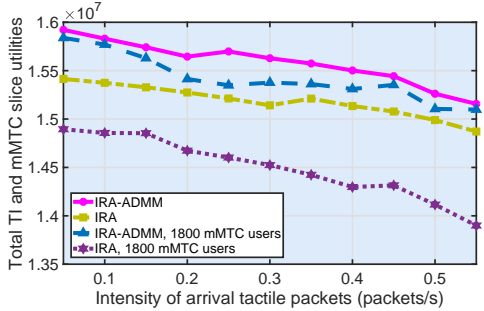}
\caption{Total TI and mMTC slice utilities vs. intensity of arrival {tactile} packets. }
\label{fig_TI_mMTC_intensity}
\end{figure}

Figure \ref{fig_TI_mMTC_intensity} illustrates the obtained total TI and mMTC slice utilities vs. arrival rates of tactile packets.
We observe that a great tactile packet arrival rate will generally lead to small total TI and mMTC slice utilities for IRA-ADMM and IRA. In fact, higher RU transmit power is consumed to deliver more tactile packets, which results in a smaller TI slice utility.
IRA-ADMM obtains greater total TI and mMTC slice utilities than IRA because IRA-ADMM is efficient than IRA in terms of bandwidth allocation.
For IRA-ADMM, it allocates bandwidth to slices in the sense of global consensus, while IRA allocates bandwidth in a local sense.

\section{Open issues}
\subsection{Edge intelligence}
Owing to the finite speed of light, the stringent responsive requirement for TI between remote human-robot pairs is difficult to be accommodated.
To address this issue, tactile edge intelligence schemes that promise real-time insights by collecting and analyzing data at the tactile edge itself must be investigated.
For example, edge intelligence enables the forecast of tactile feedback. Hence, tactile feedback can be received in real time, and then the low-latency requirement of remote human-robot communications can be satisfied without being limited by the human-robot distance.
Besides, by predictively caching tactile actuation, TI can also overcome the fundamental distance limitation of tactile applications.
However, to experience interactions with physical or virtual remote objects as intuitive and natural, accurate edge forecast results must be provided, which is challenging.

\subsection{Fast slice reconfiguration}
The dynamic characteristics (e.g., arrival and departure) of tactile traffic will hit the resource availability in TI slices.
Unfortunately, the exact dynamic characteristics are hard to be predicted.
To alleviate the effect of prediction errors on the QoS of TI slices, slice reconfiguration is desired.
Nevertheless, the slice reconfiguration indicates some higher-layer operations such as flow re-routing and VNF instance migration, which is time-consuming and needs additional resources. Besides, it may cause service interruption, which in turn degrades TI slices' QoS.
To tackle this issue, fast slice reconfiguration with a small reconfiguration cost is worthy of study.

\subsection{Private TI network}
Safety and privacy are also regarded as crucial requirements for TI.
However, the application multiplexing on a common network may result in tactile information exposure.
To avoid information exposure, the common network is required to have enhanced security capabilities as it generally takes a monolithic view towards the security architecture.
In this regard, a private TI network with complete control over every aspect of the network can be designed over the common network to transport tactile information securely and privately.
Specifically, the private TI network will be physically separate from the common network but will share partial infrastructure of the common network.
A roaming agreement can be established between the private TI network and the common network such that tactile information can traverse through the common network.
Certainly, the research on the private TI network is in its infancy, and many issues such as the network architecture, the design of flexible PHY and protocols, and the flexible spectrum sharing of the private TI network are waiting to be investigated such that optimized services can be provided.

\section{Conclusion}
An overview of the fundamental features of CoMP-enabled RAN slicing for TI was presented, followed by the TI architecture and the functional split of a CoMP-enabled RAN.
By integrating NFV and CU-DU network structure, the implementation prototype of CoMP-enabled RAN slicing for TI-involved application multiplexing was elaborated. Based on the prototype, a use case of application multiplexing was also studied.
Nevertheless, the research on CoMP-enabled RAN slicing for TI is at its nascent stage. Some key technological breakthroughs regarding security and low latency should be made before it can contribute to ``Tactile Internet" standards.


\ifCLASSOPTIONcaptionsoff
  \newpage
\fi




%
\bibliographystyle{IEEEtran}
\bibliography{CoMP_RAN_slicing}

%
\vspace{-12 mm}
\begin{IEEEbiographynophoto}{\text{ } \text{ } Peng Yang [M] (peng\_yang@sutd.edu.sg)}
is with the Singapore University of Technology and Design (SUTD). His research interests include network slicing, Tactile Internet, and UAV networks.

\textbf{Xing Xi (xixing@buaa.edu.cn)}
is currently pursuing the Ph.D. degree with Beihang university, Beijing, China. His research interests include intelligent transportation systems and UAV networks.

\textbf{Tony Q. S. Quek [F] (tony\_quek@sutd.edu.sg)}
is the Cheng Tsang Man Chair Professor with Singapore University of Technology and Design (SUTD). He also serves as the Head of ISTD Pillar, Sector Lead of the SUTD AI Program, and the Deputy Director of the SUTD-ZJU IDEA. His current research topics include wireless communications and networking, network intelligence, internet-of-things, URLLC, and big data processing.

\textbf{Hyundong Shin [F] (hshin@khu.ac.kr)}
is a Professor at the Department of Electronics and Information Convergence Engineering and the Department of Electronic Engineering, Kyung Hee University. His research interests include quantum information science, wireless communications, and machine intelligence.
\end{IEEEbiographynophoto}

%






\end{document}